BOSTON UNIVERSITY

FACULTY OF COMPUTING & DATA SCIENCES

Thesis

**HOW DO TERMS OF SERVICE INFLUENCE SOCIAL MEDIA**

**USERS DYNAMICS — A PRIVACY ANXIETY PERSPECTIVE**

by

**JINGYUAN LIU**

B.A., University of International Business and Economics, 2019

Submitted in partial fulfillment of the

requirements for the degree of

Master of Science

2026

DEDICATION

I would like to dedicate this work to my family and friends,

who supported me through the long, cold winter of Boston.

ACKNOWLEDGMENTS

I would first like to express my deepest gratitude to my advisor, Chao Su Chris, for his guidance, patience, and unconditional support throughout this project. His insightful feedback and encouragement were invaluable in shaping both the direction of this research and my growth as an independent researcher.

I am also sincerely thankful to my committee member, Professor Wesley Wildman, for his constructive criticism, and academic mentorship. His perspectives on ethics greatly enriched this work.

I would like to thank my parents for their unconditional love, trust, and support throughout my academic journey. Though once fought and misunderstood, their constant encouragement has been a source of strength, during every challenging period.

I am also deeply grateful to my friends – Cao Ren, Deng Lifeng, Xiao Yiman, Zhang Xinyu and Zhou Rui – for their constant companionship, patience, love, and comfort. During this long and challenging journey studying abroad on my own, they made it far less lonely.

This work would not have been possible without them.

**HOW DO TERMS OF SERVICE INFLUENCE SOCIAL MEDIA**

**USERS DYNAMICS — A PRIVACY ANXIETY PERSPECTIVE**

**JINGYUAN LIU**

ABSTRACT

This thesis examines how a Terms of Service update on X enabling default AI training on user content activated privacy anxiety and reshaped user behavior. Privacy anxiety is conceptualized as a structural outcome of reduced control over data use, particularly among content creators. The study finds that privacy anxiety is activated within creator communities and diffused across user groups through inter- and cross-community interaction. As anxiety escalated, engagement declined and migration intentions increased. These findings point to an unresolved dilemma in AI-driven platform governance: how user trust and autonomy can be sustained under conditions of concentrated power and data-dependent business models remains unclear.

# TABLE OF CONTENTS

# LIST OF TABLES

LIST OF FIGURES

# LIST OF EQUATIONS

LIST OF ABBREVIATIONS

| | |
|---|---|
| *CSR* | *Corporate Social Responsibility* |
| *IP* | *Intellectual Property* |
| *RAI* | *Risk Attention Index* |
| *RER* | *Risk–Nonrisk Engagement Ratio* |
| *SARF* | *Social Amplification of Risk Framework* |
| *ToS* | *Terms of Service* |
| *VSD* | *Value Sensitive Design* |

# 1. Introduction

## 1.1 Background

In October 2024, X (formerly Twitter) announced its upcoming update of Terms of Service (ToS), giving permission to machine learning training and generative AI models on user generated content. The terms explicitly state, *'analyze text and other information you provide and to otherwise provide, promote, and improve the Services, including for use with and training of our machine learning and artificial intelligence models, whether generative or another type'.* Even though AI related clause is not a new issue nowadays, this update still triggered heated public discussion and widespread opposition.

Artists, writers, scholars, and developers expressed critical concerns about user privacy and data governance. Many interpreted the change more than a contractual revision but a severe violation of consent, ownership, and creative autonomy. Consequently, the controversy quickly escalated to behavioral shifts, including decreased engagement, post deletions, and migration to alternative platforms like BlueSky.

Together, these backgrounds provide contextual background for examining the social and behavioral responses to the ToS update within a rapidly evolving, AI-driven platform environment.

## 1.2 Problem Statement

While the ToS update only authorized the use of user content for AI training on X, the resulting controversy cannot be understood as a policy issue alone.

As company increasingly turn into AI models and integrated chat systems (e.g.,

Grok), user feelings are no longer valued primarily. User participation is becoming progressively reframed as a source of data rather than a foundation for community interaction. Years ago, social media platform companies once have long preserved a relatively balanced relationship with users. However, the extensive deployment of large language models and generative AI disrupted this delicate balance. Thus, The X ToS update represents not an isolated policy change, but a critical moment of broader transformation in the reconfiguration of platform-user relations in the AI era.

Under these conditions, privacy anxiety from users emerges not as an accidental reaction, but as a structural necessity and inevitable state produced by heightened uncertainty, loss of control, and asymmetrical power relations between platforms and users.

For content creators in particular, this anxiety is rooted in the reframing of their contributions as reproducible, extractable digital labor. Concerns over IP authorship, user privacy, uncompensated data use, and long-term insecurity did not originate from the ToS update itself. Instead, the update displayed these latent tensions explicitly. Thus, privacy anxiety is utilized to be a critical lens towards subsequent reactions, functioning as an affective basic on which users interpret platform update.

Specifically, this anxiety is also not evenly distributed nor passively experienced. X's algorithmic architecture plays a central role in shaping how privacy anxiety circulates and intensifies. Recommendation systems that prioritize emotionally engaging content selectively amplify expressions of fear, anger, and distress, while marginalizing calmer, technical, or legal perspective of discussion. Under this condition, privacy anxiety is transformed from an individual affective state into a collectively reinforced and socially amplified phenomenon.

This paper explores the core issue, which is not only the controversial update to Platform X's terms of service, but also the systemic anxiety exacerbated by the platform's own architecture. Policy changes initially sparked creators' concerns about their livelihoods, such as intellectual property theft, but the platform's algorithmic incentive mechanisms further intensified this crisis. By structurally prioritizing highly stimulating content, the recommendation algorithm effectively commodifies users' anxieties, amplifying signals of distress while filtering out rational, technical analysis, thus hindering community connections and communication.

This thesis argues how the AI-driven reconstruction of platform leads to intense privacy anxiety, which is further amplified and translated into declining engagement, weakened community cohesion, and increasing user migration. Together, these patterns reveal a governance failure in which platforms prioritize AI development and engagement metrics over users' consent and the long-term stability of online communities.

### 1.3 Research Questions

To investigate the systemic collapse of trust on X following the ToS update, this thesis addresses three interconnected questions:

RQ1: The Affective Formation: How does privacy anxiety emerge following the new update, and how do users collectively interpret this policy change?

RQ2: Diffusion and Amplification Mechanism: How does privacy anxiety diffuse within and across user communities, and how do platform algorithms amplify these pre-existing affective emotions?

RQ3: Behavioral Consequence: How does this amplified anxiety shape community engagement, interaction dynamics, and user retention?

## 1.4 Contributions

This study reframes user migration not as a matter of individual preference, but as a structural outcome of platform governance under AI transformation age. By examining the link between platform policy, algorithmic design, and privacy anxiety as an affective intermediary, this research offers a systematic insight into a broader implication of privacy changes on social media ecosystems. It will contribute to discussions on digital rights, user trust, platform exit, and the future of social media, offering valuable perspectives for platform policymakers and privacy advocates.

A central contribution of this study lies in its conceptualization of privacy anxiety as a naturally arising and socially amplified condition, rather than a temporary emotional reaction. When users lose clarity over how their data is used and who benefits from it, anxiety becomes inevitable to avoid. By positioning privacy anxiety as an unavoidable byproduct of AI data extraction clause, examining how it is subsequently amplified through algorithms, this research extends existing work on privacy concerns, affective publics, and the social amplification of risk within digital environments.

Importantly, this research places a particular emphasis on the experiences of marginalized and privacy-sensitive users such as scholars and artists. Their migration may lead to the disruption of existing online communities and the loss of critical social networks. Such disruptions may further marginalize their digital identities and limit their capacity for participation in public discussion. By centering these concerns, the study aims to support the development of more equitable and inclusive platform governance models.

Finally, in the context of the AI era, where user data is increasingly treated as a resource, this research addresses the importance of corporate social responsibility (CSR) in this new era. Platform responsibility should go beyond basic legal compliance, and they should also be accountable for the social and psychological consequences of their data governance choices. Ultimately, this research contributes to a deeper understanding of the evolving dynamics between platforms and users in a rapidly shifting digital landscape.

Overall, this study shows how AI-driven policy changes activate privacy anxiety, reshape user behavior, and help explain why trust breaks down and users choose to leave. It also highlights the need for clearer corporate responsibility in the AI era.

This thesis proceeds as follows. Chapter 2 outlines the theoretical frameworks and related work. Chapter 3 introduces methods. Chapters 4 presents the activation of anxiety formation, including empirical and topic analysis. Chapter 5 shows how privacy anxiety is diffused in, across communities, and amplified through platform algorithms. Chapter 6 analyzes the behavioral consequences of these dynamics. Chapters 7–10 concludes with a discussion of implications, limitations, future directions and conclusions.

## 2. Theoretical Framework and Related Work

### 2.1 Value Sensitive Design and Misalignment

The framework of Value Sensitive Design (VSD) states that technology is not value-neutral but heavily implicated with moral and ethical values (Friedman et al., 2013). A core concern within VSD is "value tensions" or "misalignment,". Value tensions arise when platform designers must choose between competing or contradictory values embedded in the technological system, organizational constraints, or stakeholder needs (Shilton, 2013) or the values implicated in a system's design, come into conflict. (Borning, 2012)

When it comes to social media platform like X, "value misalignment" represent the logic of connectivity tends to override individual values such as privacy, autonomy, and user control (van Dijck, 2013). It also happens when incompatible logics produce structural misfits that challenge organizational legitimacy and trigger resistance among stakeholders (Kraatz, 2008).  As a result, when such value misalignments accumulate, users perceive these shifts as violations of the implicit social contract, leading to distrust, contestation, and sometimes withdrawal." (Chan, 2023)

As for ToS documents, they are not merely legal contracts but "value artifacts" that codify the platform's power to dictate how data is valued and used (Fiesler C. L., 2016). Platform values are encoded through language that expands licensing rights, legitimizes automated data scraping, and collects user content for AI model training.

The VSD framework helps explain why the ToS update triggered such a strong breakdown of trust. The conflict was not merely legal, but value-based: platform priorities around data extraction and AI development increasingly conflicted with user values such as consent, autonomy, and creative ownership. In the case of X, this value misalignment was especially obvious for content creators. The ToS update can therefore be understood as a moment when accumulated value tensions became visible, leading users to reassess the legitimacy of the platform.

## 2.2 Social Amplification of Risk Framework (SARF)

While the VSD framework explains the origin of the conflict, understanding the phenomenon further requires analyzing the mechanics of diffusion.

Originating from the seminal work of (Kasperson, 1988), SARF provides an integrative theoretical concept to explain why certain risk led to strong public concern while others do not. In the context of data privacy, SARF helps explain why simple changes in legal terms (ToS) can trigger widespread public anxiety. The framework suggests that "social stations", including the media, individuals, and groups function as amplification mechanisms that increase the volume of information, heighten the sense of risk, or reinterpret the nature of the threat (Pidgeon, 2003).

Research shows risk amplification is deeply affective. Via massive-scale experiments on Facebook, that emotional states can be transferred to others via emotional contagion, leading people to experience the same emotions without their awareness (Kramer, 2014). Tweets expressing high-arousal emotions, especially negative sentiments such as anger and fear, are also more likely to be retweeted and diffuse quickly and broadly. (Stieglitz, 2013). Within the micro-environment of a specific subculture like Art in Twitter, the shared threat to livelihood creates a "hot" emotional environment where anxiety becomes the dominant social currency, accelerating internal mobilization.

There are also transition from a localized complaint to a platform-wide crisis which scholars called "Ripple Effect" (Kasperson, 1988), a core concept of SARF, describing how risk impacts spread outward from the initial center like ripples in a pond. In social networks, this relies on structural mechanisms to bridge distinct clusters. "The strength of weak ties" is also stated while strong ties reinforce existing views, weak ties are more likely to be bridges between different groups and hence play a crucial role in the diffusion of information across social networks (Granovetter, 1973).

With fast development of modern technology, the framework has added some evolving and fundamentally changing part according to the prevalence of digital platforms, introducing the concept of "Digital Amplification (Tufekci, 2015). False news and emotionally intense content are discovered to diffuse significantly farther, faster, deeper, and more broadly than the truth in all categories of information (Vosoughi, 2018).

Within this study, SARF provides a framework for understanding how privacy-related concerns evolved from inside-community complaints into a platform-wide crisis. While the ToS update served as the initial risk signal, social interactions, community dynamics, digital infrastructures, even X itself acted as amplification stations that intensified and extended perceptions of risk. In particular, SARF helps explain how privacy anxiety diffused across communities and why emotionally charged interpretations of the policy gained disproportionate visibility and influence.

## 2.3 Related Work

Prior research has examined how users respond to changes in platform governance, particularly around privacy policies and terms of service. Studies show that sudden or opaque policy updates often generate user confusion, distrust, and resistance, especially when they affect data ownership or content rights. (Fiesler C. &., 2018) Other work has documented how privacy concerns can lead to reduced participation, self-censorship, or platform exit. (Stutzman, 2013)

Studies on privacy anxiety emphasize its role as an affective condition shaped by uncertainty, power asymmetries, and lack of control over data use. (Acquisti, 2015)More recent research on platform migration further suggests that users do not leave platforms solely due to technical features, but because of perceived value conflicts and breakdowns in trust. (Jeong, 2024).

Building on this literature, the present study connects value misalignment, affective risk amplification, and platform migration within a single analytical framework, highlighting how AI-driven governance changes can activate and scale privacy anxiety across digital communities.

## 3. Methodology

### 3.1 Methodological Rationale

This study adopts a mixed-methods research design grounded in methodological triangulation to examine the formation, amplification, and consequences of privacy anxiety following the X ToS update. The choice of this design is driven not by methodological preference, but by the multi-layered nature of the research problem itself. It spans interpretive decoding, community-level dynamics, and evolving behavioral patterns.

Privacy anxiety, as examined in this study, is neither a purely psychological state nor a directly observable behavioral outcome. Rather, it emerges as an intermediate among value interpretation, affective response, and structural amplification. No single method is sufficient to capture these strangled processes. Qualitative approaches are necessary to reveal how users' own interpretation and concerns, while quantitative approaches, in turn, are required to assess how these interpretations function across communities, vary over time, and become unevenly amplified.

Accordingly, this study combines qualitative netnography with quantitative metric-based analysis, treating them as complementary rather than hierarchical. Netnography provides insight on how the policy is decoded and how privacy-related concerns are narratively framed. Quantitative analysis then operationalizes these insights into measurable indicators, enabling the examination of large-scale patterns in anxiety expression and engagement intensity across communities.

Together, this mixed-methods triangulation enables a more comprehensive and internally coherent analysis than any single method could provide. This methodological rationale lays a foundation of the research design detailed in the following sections.

3.2 Research Design Overview

3.2.1 Value Framework

To systematically analyze the value between the platform's governance policy and user discourse, this study adopts a value framework. Value is conceptualized as an attribute to describe platform governance and a justification for legitimizing certain platform clauses (Chan, 2023). This framework enables systematic comparison of how values are articulated at the level of both governance policies and interpreted within user discourse.

The value framework is informed by established platform studies literature, including engagement and authenticity (Hallinan B, 2022), privacy (R, 2019), fairness (Van Dijck J, 2018), while being adapted to the specific context of recent generative AI governance. Through cross-referencing these theoretical perspectives with X's General Rules (X.), our study identified Intellectual Property (IP) as a distinct and analytically central value dimension. (X).

Since the new ToS update directing reconfigure ownership, authorship and data use rights, Intellectual Property (IP) is identified as the primary focus of the value misalignment in our study. Unlike other values that function as background governance principles, IP-related values are explicitly contested and discussed in both policy language and user responses. Recognizing this asymmetry, the value framework distinguishes IP from other values.

A detailed codebook and operational definitions of X's platform values are provided in the supplementary material.

### 3.2.2. Temporal Design

The research adopts an event-driven temporal design structured around two key policy time associated with the updated X ToS. The first milestone is the public announcement of the new ToS on October 16, 2024, followed by its formal deployment on November 16, 2024. These two points mark distinct phases in how users encounter, interpret, and respond to the policy change.

To capture the dynamics of user discourse surrounding these events, the study defines a three-stage temporal window including the month preceding the announcement, the immediate reaction period following the announcement, and the post-deployment phase. This temporal segmentation defines the analytical windows used throughout the study. See Table 1 for details.

| Time period (30days) | Window Name | Description |
|---|---|---|
| 9.15–10.15 | Pre-Announcement | Baseline period for normal discourse |
| 10.16–11.15 | Announcement Reaction | Immediate reaction and outcome |
| 11.16–12.15 | Post-Announcement | Sustained behavioral changes |

Table 1: Time Period Segmentation

### 3.2.3 Analytical Component

Based on the mix-method triangle framework, interpretive and measurement components are utilized.

The interpretive component focuses on how users collectively understand the ToS update, consisting of construction of a community-based taxonomy and netnography observation to decode privacy concerns and value among user groups. The measurement component then examines how these interpretations scale and circulates at a larger level. Using a set of quantitative metrics, including anxiety score, topic value, engagement, Risk Attention Index (RAI), Risk–Nonrisk Engagement Ratio (RER), it captures broader patterns across communities and over time.

Netnography analysis, along with anxiety-related metrics, answer RQ1 by examining how privacy anxiety is framed following the ToS update. RQ2 is then discussed through community-based engagement metrics that capture structural differences in risk amplification across groups. RQ3 is examined through behavioral consequences including engagement and user retention.

### 3.3 Qualitative Methods

*3.3.1 Taxonomy Construction*

User discourse was first organized through a community taxonomy to define the primary units of qualitative analysis. Related posts were retrieved using a  keyword-based search strategy following the announcement on October 16th, 2024. Keywords included 'Terms of Service', 'ToS', 'Privacy', and 'AI Modeling'.

Based on posing patterns and topic relevance, accounts were grouped into three distinguished categories: privacy-sensitive communities**,** non-sensitive communities**, and** news media accounts. Table 2 summarizes the conceptual characteristics of each category and their analytical role in the study. A full list of accounts is provided in Appendix A.

| Community Category | User types | Characteristics | Role |
|---|---|---|---|
| Privacy Sensitive Communities | Creator, Developer, Scholars | Users whose posts frequently reference authorship, copyright, or data use in context of AI and platform policy. | Analytical focus |
| Non-Sensitive Communities | Sports, F1 drivers | Users active during the same period with no discourse on privacy or AI related issues | Baseline comparison group |
| News Media Accounts | News Media | Accounts primarily engaged in reporting or reposting policy updates with limited personal commentary | Information intermediaries |

Table 2. Classification of Communities and example users.

### 3.3.2 Netnography Analysis

Based on the group taxonomy, the study further adopts an observational netnography method to understand how the privacy sensitive communities decoded the updated ToS through ongoing posts. It focuses on recurring insider community language and emotional narrative by users when articulating privacy-related concerns against the abbreviated law policy. (Kozinets, 2010)

Through iterative reading of posts and related replies and quotes, the analysis identified shared interpretive patterns into a community level. This process also generated an empirically grounded anxiety lexicon, which was subsequently used to quantify shift in anxiety expression. The structure of the lexicon is summarized in Table 3, with full details provided in appendix A.

| Category | Subtype | Example Terms |
|---|---|---|
| Direct Anxiety | Risk, fear, harm | scrape, exploit, fear, unsafe |
| Metaphorical Anxiety | Data misuse | Losing control, stolen data, not protected |

Table 3: Structure of anxiety lexicon

## 3.4 Quantitative Methods

To systematically examine patterns identified in the qualitative analysis, we constructed two distinct datasets: a corpus of platform policy documents and a longitudinal corpus of user discourse. Five metrics are utilized to analyze them.

### 3.4.1 Data Collection

To examine large-scale patterns associated with the ToS update, two datasets were constructed: a corpus of platform governance documents and a longitudinal corpus of user discourse.

The policy corpus consists of three versions of X's official Terms of Service, released in May 2023, August 2023, and October 2024. They serve as reference materials for comparative analysis of platform value over time.

The user corpus was collected via the X API based on the community taxonomy. Collected metadata include text, timestamp, user ID and interaction counts (replies, reposts, quotes, and likes). They were structured according to the temporal design, with all observations aligned to the pre-announcement, announcement reaction, and post-announcement phases. Summary statistics by community and time period are reported in Table 4.

| Community | Pre (#posts) | During (#posts) | Post (#posts) | Total |
|-----------|--------------|-----------------|---------------|-------|
| Sensitive | 2038 | 3336 | 2148 | 7522 |
| Non-sensitive | 99 | 120 | 105 | 315 |

Table 4: Number of Posts by Community and Phase

*3.4.2 Metrics*

To quantify users' reactions to the ToS update, we construct a set of metrics that operationalize anxiety expression, value orientation, engagement intensity, and privacy risk attention. Full technical specifications are provided in the Appendix C.

A. Anxiety Score

The Anxiety Score measures the intensity of privacy-related anxiety expressed in posts. A hybrid approach combining NLTK Anxiety lexicon and prompted GPT-4o-mini are utilized to generate the anxiety score.

B. Topic Value

Value classification is operationalized according to the value codebook and applied to platform policy and user discourse. Using sentence-level labeling, texts were assigned to predefined value categories. See supplementary material for detailed value for each sentence of ToS.

C. Engagement

Engagement captures interaction intensity based on replies, reposts, and quotes. It serves as both a level of engagement across community (Davies, 2024), and an indicator of affective amplification (used in Chapter 4).

D. Risk Attention Index (RAI)

The Risk Attention Index (RAI) measures the proportion of community discourse to privacy-related risk. It captures how much a community focuses on privacy risk compared to overall conversation volume. A higher RAI indicates a community with strong attention to privacy threats.

E. Risk–Nonrisk Engagement Ratio (RER)

The Risk–Nonrisk Engagement Ratio (RER) compares engagement levels between risk-related and non-risk-related content.

An RER greater than 1 indicates disproportionate amplification of risk-related discourse, suggesting heightened emotional sensitivity with privacy anxiety. And an RER below 1 suggests no heightened engagement with privacy-related content, indicating lower affective reactivity.

Together, these mixed methods jointly enable a comprehensive, coherent framework to address the research questions.

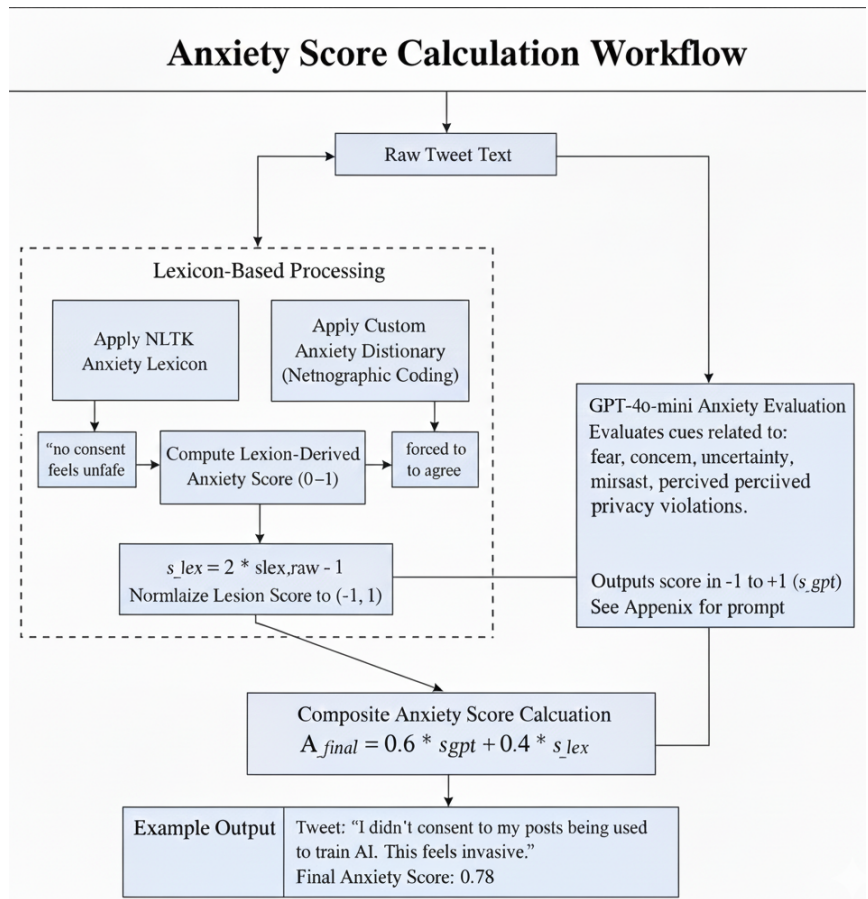

Figure 1: Anxiety Score Calculation

## 4. The Activation of Privacy Anxiety

Based on an overall Social Amplification of Risk Framework, this chapter studies the starter stage of the risk process: how a platform governance change is interpreted as the risk signal in the first place.

Specifically, we conceptualized privacy anxiety as a value-induced risk signal, whose activation is explained through a Value Sensitive Design (VSD) perspective. By tracking shifts in platform values and corresponding user interpretations, this chapter addresses RQ1 by showing how the X ToS update was decoded and privacy anxiety is activated.

### 4.1 Value Misalignment: Precondition of Risk

In SARF, risk signals do not arise from objective hazards alone, but from social processes that render certain conditions meaningful and threatening. In platform governance contexts, these processes are laid on inherent value. This section therefore begins by examining value misalignment between the platform and its users as the structural precondition for risk signal formation.

From a VSD perspective, technologies and policies encode particular value priorities. When these priorities diverge from user expectations, latent tensions can become activated. In the case of X, the ToS update did not introduce entirely new concerns, but made existing value conflicts visible, creating the conditions under which privacy anxiety could be triggered.

*4.1.1 Value Shift in Platform ToS*

We first examine the platform shift of values by tracing evolution of ToS over time. Based on our value codebook, the normalized frequency of topic value from May 2023 to October 2024 is visualized in Figure 2.

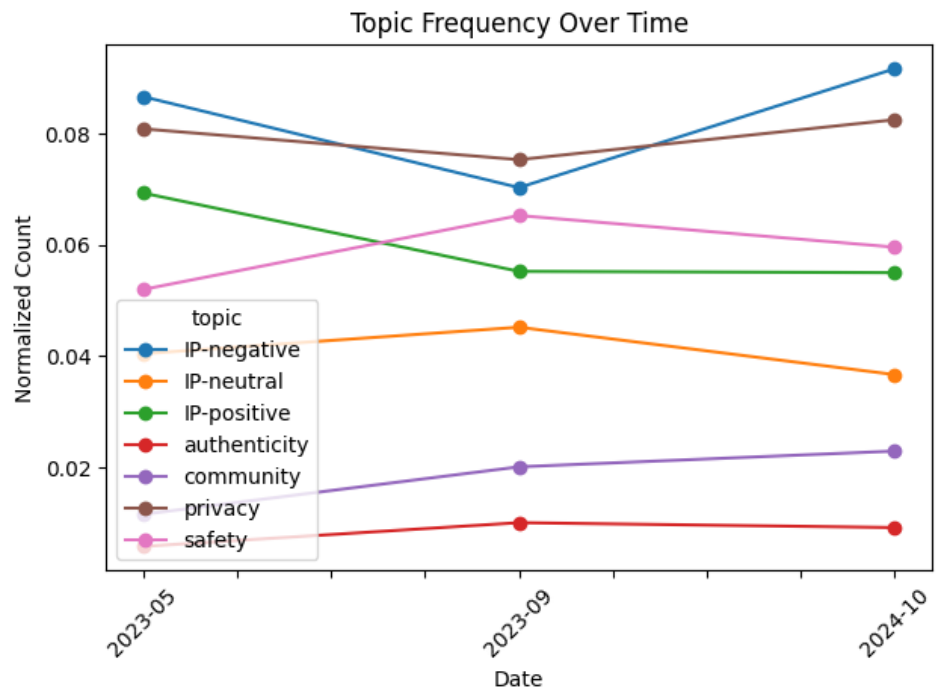

Figure 2: Value Frequency of ToS Over Time

The figure indicates a structural shift in IP-related value: a transition from a relatively diverse value distribute (May 2023) to one dominated by IP-negative (October 2024). A "scissor-like" pattern can be easily discovered regarding IP discourse topic. Specifically, IP-positive content, associated with user ownership and data protection consistently dropped by 20% by 2024-10. While IP-negative content, emphasizing data extraction surged aggressively and becomes dominant.

This unique pattern indicates beyond a mere change in legal language, but a reorientation of platform values, switching from user rights to corporate control. From a VSD perspective, this shift constitutes a reconfiguration of the platform's value commitments, laying the groundwork for user's distrust.

*4.1.2 Divergence in User Interests*

As for users' reactions, users are divided into two groups based in the qualitative taxonomy: a non-sensitive baseline group, and a privacy-sensitive target group.

The analysis of their post topic reveals a fundamental divergence in value orientation. The non-sensitive group preserves a stable topic pattern, with negligible attention to privacy or IP-related issues (Figure 3). On the contrary, the sensitive group exhibits strong focus and surges on privacy-related discourse (Figure 4).

These surges (green line) also temporally aligned with key moments about the ToS announcement and deployment, indicating the attention to value-relevant issues is event-driven rather than habitual. Being hyper-vigilant to related topics, the users from sensitive group share a sense of constant alertness, distinct from the passive compliance of non-sensitive users.

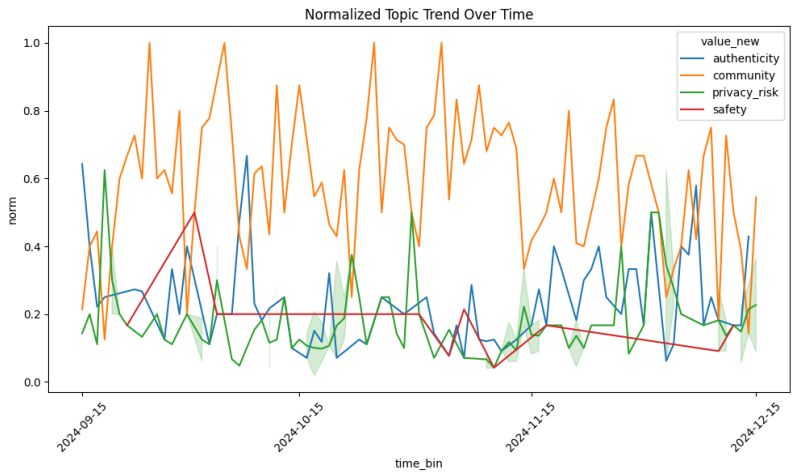

Figure 3: Topic Trend from Sensitive Group

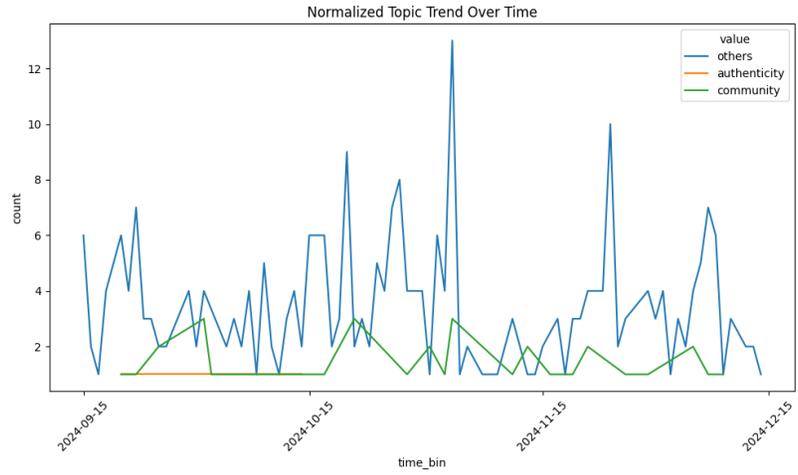

Figure 4: Topic Trend from Non-Sensitive Group

## 4.2 Interpretation of the New ToS

Value misalignment alone does not produce risk. It must be interpreted as such by affected communities. Beyond topic prevalence, privacy anxiety is also shaped by interpretive process. This section then observes the interpretive language of policy change through netnography.

Netnography analysis of highly engaged posts reveals that the ToS update is widely interpreted more than just a contractual adjustment but as a realignment of value and betrayal. Users frequently frame the new ToS as a legalization of theft, and loss of IP ownership. As one user states, *"You no longer hold exclusive rights... X can use, share and profit off your original work."* References to regulatory frameworks like GDPR and the EU AI Act further grounded these interpretations in legal and ethical discourse. Highlighting jurisdictional ambiguity and absence of clear opt-out mechanism, they recognized the AI clause as a deeper moral degradation using expressions like 'betrayal', 'exploitation', and 'decay'.

Through these specific languages, users frame the policy change both as invasion of users' values and betrayal of trust. Together, this collectively formed an actionable threat narrative, leading to a final risky interpretation.

## 4.3 Privacy Anxiety as Risk Signal

Bringing together platform value shifts and user interpretations, this section conceptualizes privacy anxiety as an affective form of the risk signal generated through value misalignment, answering RQ1.

With a dual stakeholders view on both platform articulation and users' orientations, privacy anxiety does not emerge in a vacuum. Rather, it is the product of a misfit between person-organization relationships. Drawing on person–organization fit theory (Kristof, 1996), this gap refers to the growing divergence between what the platform legitimizes and what users perceive as acceptable.

With an increasing shift towards IP-negative value framing, privacy-sensitive users place greater emphasis on authorship and privacy-oriented value. For creators and artists, the new ToS is not merely a policy update but a symbolic violation of creative dignity. Within this gap, in SARF terms, privacy anxiety constitutes a risk signal. It translates value conflict into a socially recognizable form that can be communicated, shared, and, in subsequent stages, amplified. This chapter therefore establishes the foundation for the amplification dynamics examined in the next chapter, where this risk signal moves beyond formation into circulation.

## 5. The Amplification of Anxiety

Based on SARF, a social media platform functions not merely as a transmission channel, but as a complex web of 'social stations' where risk signals are decoded, processed, and recoded. Having established how privacy anxiety forms through value misalignment and collective interpretation, this chapter examines how that anxiety is subsequently amplified within the platform, answering RQ2.

### 5.1 Amplification Across Interaction Levels

This section examines how privacy anxiety is amplified within interaction networks, focusing on how risk signals are decoded and transformed through user engagement.

### *5.1.1 Interaction-Level*

We first examine amplification at the level of individual interaction, starting with signal decoding. Figure 5 compares the distribution of Anxiety Scores between privacy-risk posts (privacy- and IP-negative) and non-risk content. The privacy risk posts exhibit a systematically higher anxiety level, with significant rise in mean and the minimum value, while non-risk content shows a neutral median. This pattern indicates that users read the signal through an anxiety affective way, gradually accumulating tension.

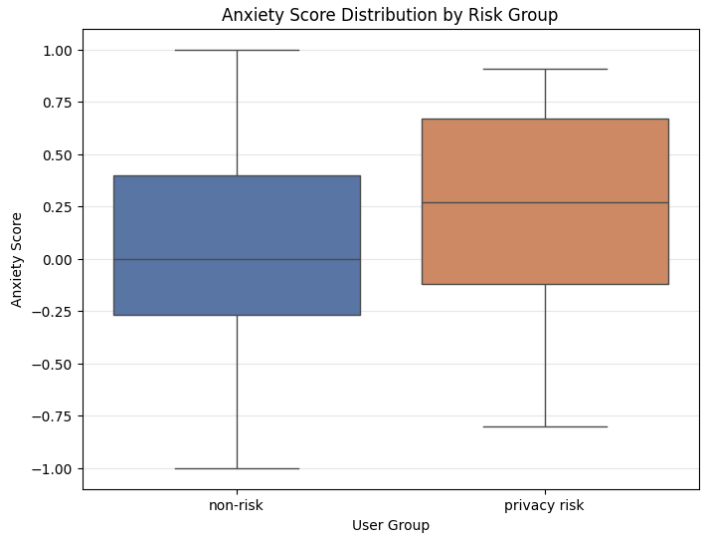

Figure 5: Anxiety Distribution by Risk Group

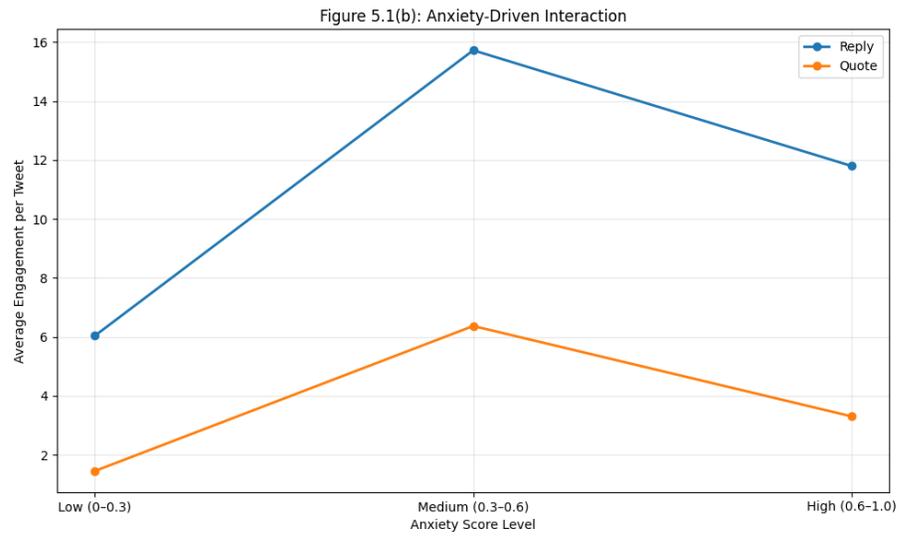

Figure 6: Interaction Categorized by Different Anxiety Level

Building on this baseline, a necessary shift of role from a receiver to an amplifier through communication is required to achieve the diffusion of risk signal (Pidgeon, 2003). To testify this mechanism, the relationship between anxiety score and initiative engagement, measured by quotes and replies is examined (Figure 6). A clear non-linear, inverted U-shaped pattern is revealed.

At low anxiety levels (0–0.3), engagement remains minimal, with low-anxiety content failing to cross the activation threshold. Engagement peaks at moderate anxiety levels (0.3–0.6), increasing nearly fourfold compared to the baseline level. While further increases (0.6–1.0) in anxiety shows a structural decay in engagement, suggesting an inhibitory effect.

This pattern indicates that anxiety functions as a conditional driver of amplification rather than a linear force. Signals encoding moderate to high anxiety, possessing sufficient emotional intensity to attract attention while maintaining enough social acceptability to maximize broad resonance without triggering defensive rejection.

At an aggregate level system level, these interaction dynamics scale into a system-level outcome, a Risk–Nonrisk Engagement Ratio (RER) of 3.77. A significant amplification of privacy risk is shown, with risk-related posts received nearly four times as much engagement as non-risk posts, going viral across the X network.

This imbalance ratio provides quantitative evidence how micro-level affective engagement accumulates into a measurable system-wide amplification effect.

*5.1.2 Cross-Community Amplification*

While system-level amplification reflects aggregate engagement, amplification is unevenly distributed across communities.

Different communities possess a structural heterogeneity when it comes to sensitivity to privacy anxiety and amplification effect of risk. Figure 7 maps four communities (creators, scholars, developers, news media) into a risk landscape with Risk Attention Index (RAI) and RER, level of attention to risk in daily posting and extent of amplification level.

Creators occupy a high-attention, high-amplification region. They have the highest sustained attention (RAI ≈ 0.046) and extremely high response intensity (RER ≈ 20). Their personal preference of risk topics eliminates the distance between the information and recipients, leading to an extremely strong dissemination effect. Relatively, news media has a low sustained attention but exceptionally high amplification effect. With an RER of 30.0, a single news report on a relevant privacy risk topic driven from ToS can trigger massive response.

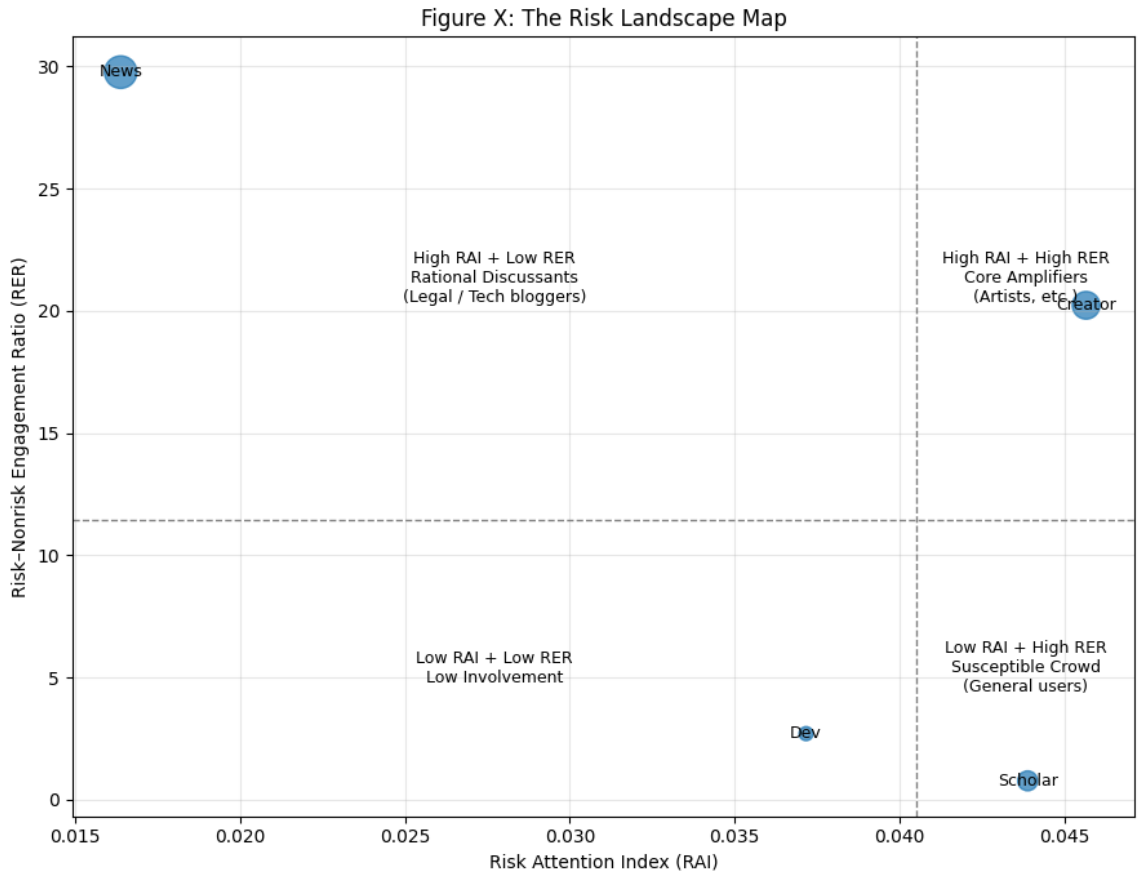

Figure 7: Risk Landscape Map

On the contrary, scholars and developers exhibit relatively high attention with limited amplification. With relatively similar interest in the topic compared to content creators, their response is very limited.

Together, these patterns reveal that communities differ not only in their sensitivity to privacy risk, but also in how attention is transformed into engagement. Amplification is therefore structurally differentiated rather than uniformly distributed.

Cross-community interaction further shapes this process. The interaction network graph (Figure 8) reveals a critical distinctive trait in the ecosystem: Isolation. While cross-community exchange remains virtually non-existent, an overwhelming dominance of self-loops within communities stands out. The amplification exhibits a relatively parallel and internally reinforced amplification mode, rather than chain-like transmission. As a result, high-anxiety interpretation remains to be unsolved and substantially recalibrated within communities.

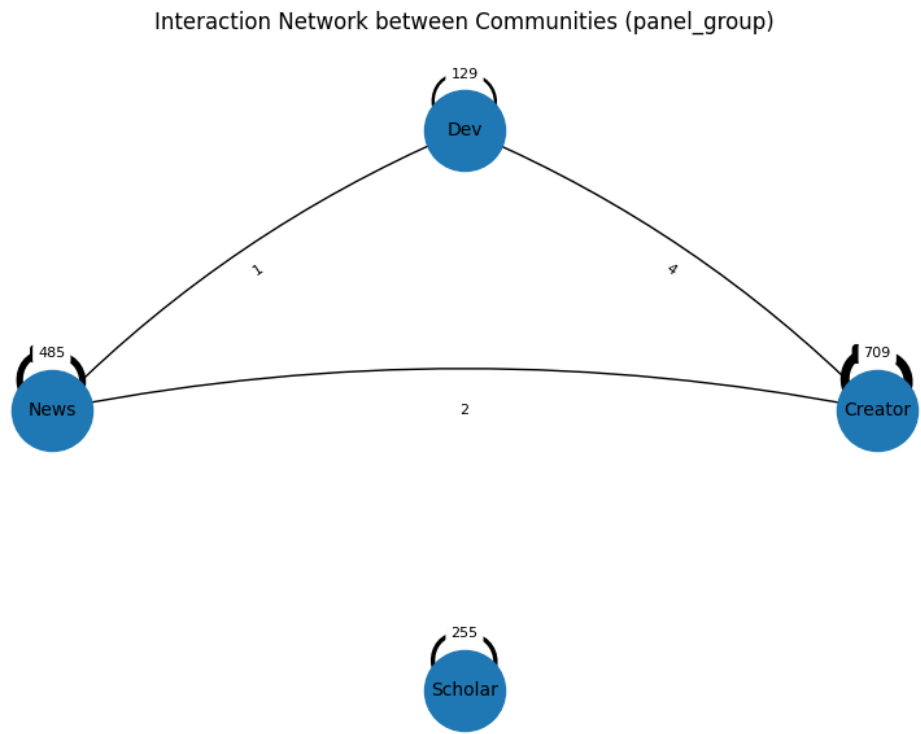

Figure 8: Interactive Networks

For different stations, they act and amplify the same privacy risk, simultaneously but independently, each driven by its own internal affective logic, discursive norms, and community-level value alignment.

## 5.2 Digital Amplification

While previous section analyzed through communities' level of risk propagation, amplification also evolves in the digital age with a technological infrastructure. To reflect on this, we examine how engagement-related signals are associated with content visibility. Figure 9 visualizes a linear regression analysis examining the relationship between Anxiety Score (input) and Log-Engagement (output/visibility). This analysis, somehow, exposes the platform's reward mechanism in an observational lens。

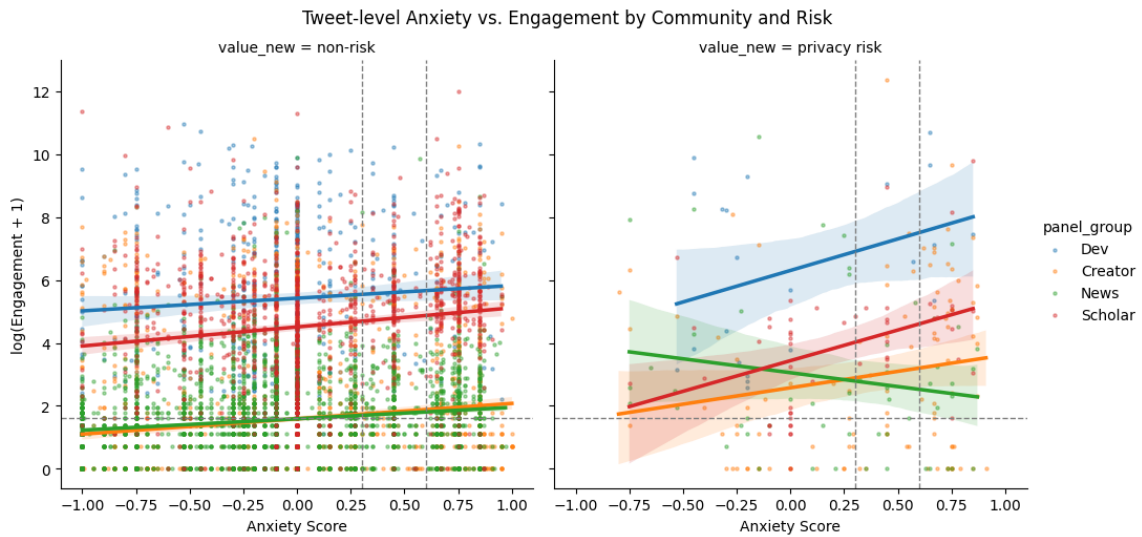

Figure 9: Anxiety & Engagement

First, a structural difference between risk and non-risk posts is discovered. For non-risk posts, the regression slopes remain relatively flat across communities, indicating that higher anxiety on everyday topics does not significantly associate with increased engagement. However, when it comes to privacy risk related discussions, a steeper positive slope of the regression is shown. This indicates that anxiety is positively and largely correlated with engagement on and only on privacy-related risk discourse.

Notably, the strength of this association varies across communities. The steepest slopes appear within developer and scholar's communities, where interactions volume remain little on low-anxiety posts, but skyrockets as anxiety score increases. Creator communities exhibit a similar but weaker pattern, indicating a strong selective preference on engagement across non-risk and privacy-risk posts at an outcome level.

Together, the regression results demonstrate that anxiety expression is differentially distributed with engagement outcomes depending on content types. Engagement remains largely insensitive to anxiety in non-risk contexts, while privacy-risk posts show a strong positive association between.

## 6. Behavior Outcome

With previous chapter explaining how interactive networks, social structures, and algorithmic mechanisms effectively amplify privacy anxiety, the study then comes to examination on how privacy anxiety are translated into observable behavioral outcomes. In SARF, the final stage of the amplification process is the "ripple effect," preserved as behavioral changes, particularly how individuals respond to perceived threats, thereby altering their digital lifestyles.

Rather than assuming a single terminal outcome, this chapter operationalizes behavioral consequences through two empirical proxies: changing engagement patterns over time, and migration intention with partial realization.

### 6.1 Engagement as a Behavioral Proxy

While Chapter 5 documented heightened activity during the amplification phase, a state of structural fatigue of engagement is followed. As shown in Figure 10, the relationship between time and daily posts count shows a typical burnout pattern in the creator community (blue line). In early November, activity volume surges extremely high, achieving approximately 140 posts per day. However, after policy implementation on November 15th, activity declines significantly, falling below pre-announcement baseline levels (September–October).

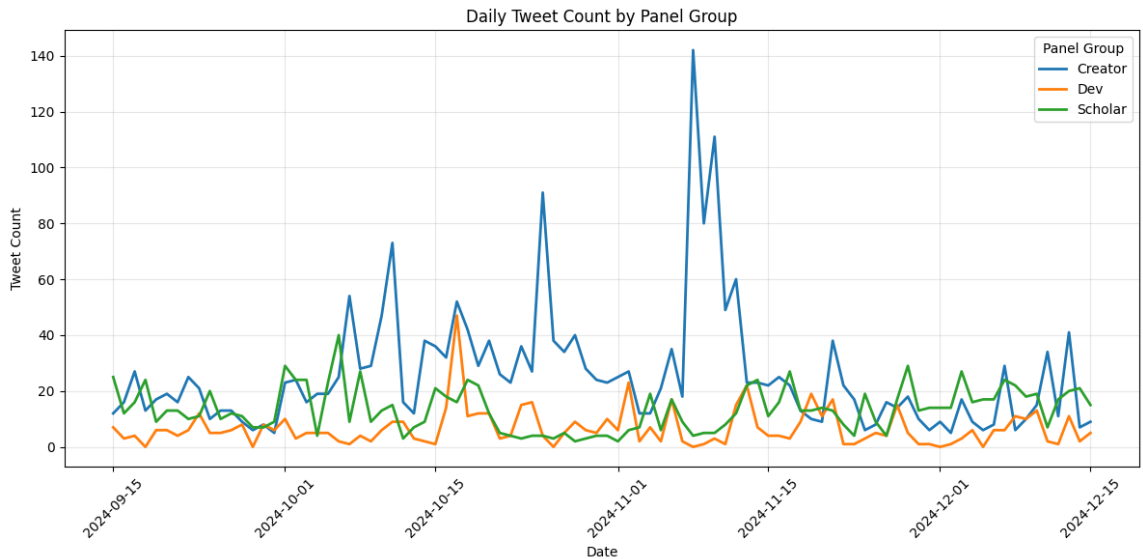

Figure 10: Daily Tweet Count

Importantly, this decline is not merely a return to equilibrium. Both the range and maximum of daily activity decrease, indicating reduced participation intensity. While the developer and scholar communities maintained a relatively stable baseline level, the creator community showed signs of withdrawal after the brunt of privacy anxiety.

These patterns suggest a behavioral withdrawal within the most affected communities, creators. Instead of dynamic fluctuation, strategic silence and permanently reduced user participation is depicted after burnout of privacy anxiety.

### 6.2 Migration Intention

While reduced engagement reflects short-term withdrawal, a more explicit behavioral response to privacy anxiety is users expressed intention to leave the platform. To distinguish actual behavioral outcome, migration intention and subsequent posting behavior are discussed separately.

To capture migration intention, we analyzed users who explicitly expressed plans to leave the platform, using keywords such as "migrating," "leaving," and the names of competitors like BlueSky in their posts and updated introduction.

As shown in Figure 11, the creator community accounts for approximately 75% of all migration-related posts, while developers and each contribute around 12.5%. Figure 12 further shows, despite the widespread dissatisfaction, creators exhibit the highest within-group migration intention. This evenly distribution algins with earlier findings on highest levels of anxiety and amplification within creator communities.

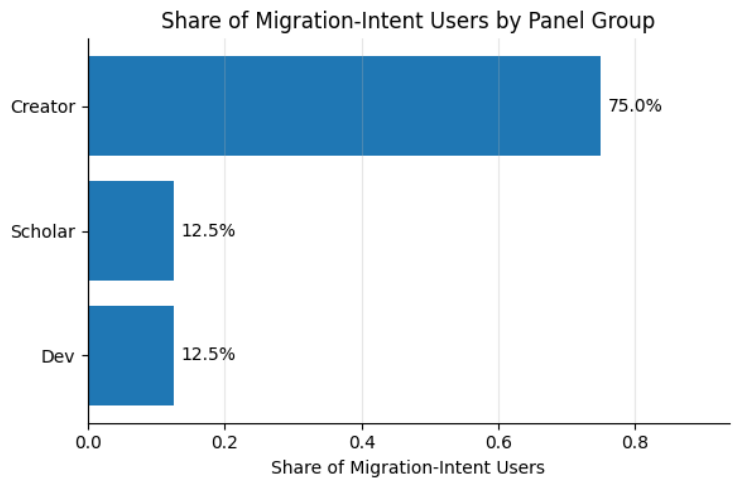

Figure 11: Migration distribution

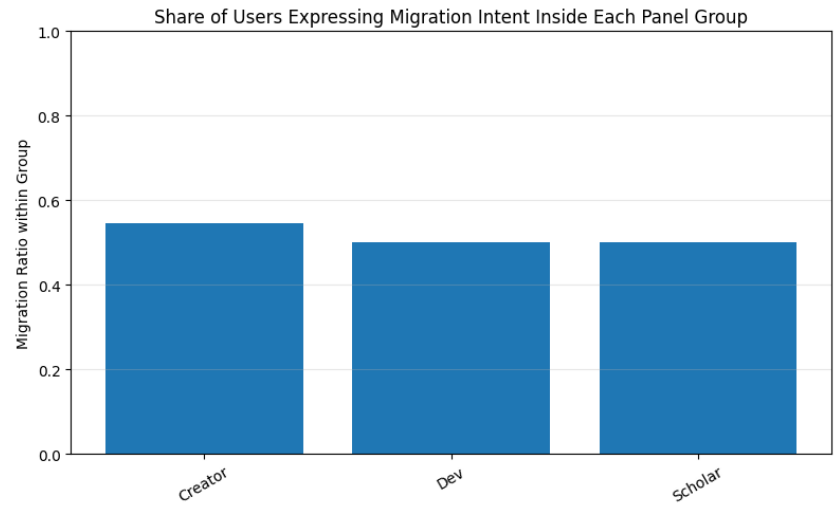

Figure 12: Migration User Proportion

However, expressed intention does not directly translate into actual migration. To assess behavioral realization, we tracked posting activity following migration-related posts. Only approximately 20% of users who explicitly announced migration ceased posting entirely on X after their migration-related post, indicating realized departure. The remaining majority, however, either reduced activity temporarily or resumed posting after a short while.

In summary, the behavioral consequences of the ToS update serves less as an immediate exit and more as a contested behavioral adjustment. For many users, migration intention serves as an exploratory response to anxiety rather than a final decision. Nonetheless, even partial realization indicates a measurable erosion of platform attachment.

7. Discussion

This study shows the undergoing mechanism of a privacy crisis observed on X following the ToS update. It was not triggered by external attack or misinformation, but emerged from internal, structural conditions shaped by the platform itself. By prioritizing AI training and engagement metrics over user trust and community stability, X created a self-reinforcing cycle where privacy anxiety was activated, amplified, and translated into mutual disengagement and migration.

7.1 Interpreting Privacy Risk

Addressing the first research question, the findings show that privacy anxiety did not originate with the ToS update itself but was activated by it. Constant concerns among content creators regarding IP, consent had already existed but not exploded before the policy change. The ToS update rendered these concerns explicit and unavoidable, triggering privacy anxiety as a structural. From a Value Sensitive Design (VSD) perspective, this reveals a misalignment between platform values and user values: creators' expectations of ownership, and respect were no longer reflected in platform governance decisions.

Then, risk is first interpreted through an affective lens, establishing a heightened anxiety baseline. Then, anxiety functions as a conditional driver of engagement, producing a stage-based and dynamic amplification pattern at the individual level. However, the meaning of that signal was not fixed. Instead, it was decoded, framed, and circulated differently across communities.

For creators, they are the engines driving the continuous spread of anxiety,

keeping the topic visible and emotionally charged. Updates of ToS are not an abstract legal issue, but a direct threat to their daily behaviors, understanding as an intellectual property theft. This personal relevance eliminates the distance between the information and the recipients, leading to an extremely strong dissemination effect.

News media has a unique structural position in this phenomenon, acting as information amplifiers. As a bridge for risk communication, transforming privacy concern from a niche sensitive group into a widespread risk across the platform, news media further amplifies the localized anxiety of content creators and spread it to the public.

By contrast, scholars and developers exhibited patterns consistent with decay. Due to rational professional norms and detached analysis rather than emotional dissemination, their ability to retransmit risk signals is constrained. With their only technical rationality apart from emotional impact, preventing them from falling into a panic circle, thus only receiving and processing the information but do not amplifying its threat that much.

Together, risk amplification was not uniform across users. Instead, it emerged from the interaction between community-specific interpretations and role-based norms, determining how signals were expressed and retransmitted, partially answering RQ2.

7.2 Amplification within the system

Apart from structurally patterns across communities, platform-level incentives also act an undeniable role in amplification: community isolation and algorithmic preference.

Network analysis reveals undeniable evidence of echo chamber effect where anxiety is repeated and reinforces within homogenous communities without internal validation. With the system lacking a corrective feedback loop that the SARF assumes to be essential for recalibrating public risk perception, which turned out to be scholars' and developers' rational posts, the X systems itself is unable to moderate anxiety within the creator community.

This isolation was further reinforced by the platform's recommendation environment. While interaction metrics fundamentally reflect human responses, they also act as the currency of visibility in the recommendation ecosystem.

Through regression analysis, we discover through actual performance that algorithms on X are not neutral tools, but functioning as active amplification stations that select, filter, and boost content based on engagement probability. In SARF terms, the platform itself functioned as an amplification station, reshaping risk signals rather than neutrally transmitting them.

X's recommendation system systematically rewarded anxiety-intense privacy-risk content with visibility, while filtering out calm, technical ones. Technical explanations is systematically filtered out unless it is accompanied by visible anxiety. In effect, the platform imposes a visibility tax with emotional restraint. As a result, the broader public is exposed to a distorted image that the tech community is in a state of panic, simply because their calmer insights are algorithmically invisible.

In this case, anxiety is no longer just an emotion, but has become a high-value, detectable signal that the algorithm prioritizes for distribution. The combined effect of isolation and algorithmic mediation produced amplification without correction. By systematically rewarding content that arises strong emotional responses disproportionately, the platform's design effectively transforms the policy update into widespread, large-scale panic, cruelly exploiting user to gain traffic, transforming policy disasters into short-term surges in engagement, exploiting users who are threatened and betrayed on privacy issues,

7.3 Corporate Governance and Strategic Indifference

Chapter 6 explores the aftermath of this tsunami of privacy anxiety. As the platform gradually transforms in users' perception from a home for community to a selfish environment used for data collection, users are actively seeking out alternative platforms such as BlueSky, which market themselves as safe places. This leads to the forthcoming decline in engagement and growing migration intention, answering RQ3. However, this migration mostly remains temporary and not behavioral.

The dynamics then raise broader questions about platform governance and corporate responsibility in the AI era. Rather than framing the ToS update as a policy communication failure, this study suggests that the resulting crisis reflects a deeper successful governance logic.

In the short term, the amplification of privacy anxiety generated a substantial surge in engagement without requiring additional governance action. Beyond short-term metrics, it also functioned as a mechanism of user filtering.

As anxiety intensified, some creators disengaged or temporarily exited, while a subset remained or eventually returned. Due to the platform's market dominance and concentration of audience attention, departure was rarely final. Many users who resisted the policy fought and failed, gradually forced to accept data extraction as a normalized and necessary condition for participating in the platform.

Over time, this dynamic would enable a subtle but consequential cognitive shift that creative activity is reframed as data production, normalizing extraction without generating corresponding legal risk or accountability for the platform. Rather than persuading users, the platform utilized structural constraints and engagement incentives to do the work. In this sense, the crisis did not merely disrupt the community. It actively reshaped it.

In this configuration, there is no need for the platform to suppress disagreement or force users to agree. Every act of users' debate, frustration, withdrawal, or return contributes to engagement, data generation, and ultimately consolidates a more manageable user base. Therefore, what appears to be depressing resistance from the user's perspective actually serves to strengthen and stabilize the platform from the company's perspective.

## 8. Limitations

This study has several limitations.

First, data collection is strictly constrained by X's API access and scraping mechanisms. Without a pro-premium API access which costs $5000 per month, keyword-based searches are limited to posts only from the most recent 30 days. Older posts related to privacy and ToS discussions can only be retrieved through user-based queries, making large-scale historical data collection difficult. In addition, the platform only provides interaction counts but not the full text of all replies and reposts. It further jeopardizes the richness of our analysis. Count of retrievable historical posts per user is also restricted to make it difficult to reconstruct long-term activity for highly active users.

Second, this study focuses on a specific time window following the ToS update. As a result, it cannot fully capture how the platform ecosystem evolved in the longer term, including whether privacy anxiety persisted, faded, or transformed over time.

Third, while the study identifies patterns of anxiety diffusion and algorithmic amplification, the analysis remains relatively indirect. Due to data and platform constraints, it cannot observe the recommendation process directly or fully disentangle algorithmic effects from user-driven engagement dynamics.

## 9. Future Work

These limitations suggest several directions for future research.

First, future studies could incorporate larger and longer-term datasets, either through expanded API access or archived data sources. This would allow for a more complete reconstruction of privacy-related discourse and engagement patterns over time.

Second, how the platform ecosystem evolved one year after the ToS update could be further examined. This includes assessing whether anxiety-related dynamics stabilized or shifted, how remaining communities adapted, and how users who migrated reshaped the ecosystems of alternative platforms.

Finally, future research could develop more fine-grained analyses of algorithmic amplification. By modeling visibility, timing, and exposure in greater detail, future studies could better capture how algorithms interact with user anxiety and how amplification mechanisms operate.

## 10. Conclusions

This study examines a platform-driven governance decision by X's ToS update that reshaped community dynamics.

The findings show that privacy anxiety was not uniformly experienced across users but was disproportionately concentrated among content creators whose labor is both highly visible and highly extractable. Through interaction networks and algorithmic mediation, this anxiety was amplified and self-cycled. Over time, heightened anxiety translated into declining engagement and measurable intention but limited actual behavioral migration.

Importantly, these outcomes should not be understood solely as a failure of corporate management. Instead, they raise broader questions about corporate governance and responsibility in the age of generative AI. The case of X illustrates a tension between data-driven platform strategies and social foundations that sustain online communities. More critically, the event exposed a persistent imbalance of power between platforms and users. They reveal a structural condition in which meaningful exit is constrained by monopoly power and possible concentration of visibility, under which resistance often remains symbolic, while compliance emerges gradually through fatigue, dependence, or lack of competitive alternatives.

Rather than demonstrating the success or failure of a particular platform strategy, this case points to a broader structural problem: how balance can be maintained between corporate power and user autonomy within highly concentrated digital ecosystems. Current governance frameworks offer limited mechanisms for addressing affective harm, value erosion, or asymmetries of dependency. How such imbalances might be corrected, or whether they can be corrected at all, remains an open question, warranting further theoretical, empirical, and regulatory investigation.

## APPENDIX

### Appendix A: Details on Data Collection

A.1 Users Accounts:

News_handles:

['80Level','RhinozzCode','YourAnonNews','LizCrokin','DMichaelTripi','RobertBohan']

Scolar_handles: ["LuizaJarovsky", 'BrianEMcGrath', 'BrianRoemmele', 'OAlexanderDK', 'UK_Daniel_Card']

Creater_handles:['celiander','BigEsqBae_','astratria','CiblesGD','aInstsua','rainymegane','dyarikku', 'dreamosaurus','lemonjamdraws','DriftingEmbers', 'LitheNightshade', 'Goblin_Alchemi', 'kortizart','BCD3258','RickyNyanheart']

Dev_handles: ['PirateSoftware','RiseNGrindGame']

From September 15th to December 16th, a total of 29 X users with 7522 tweets are identified. After excluding 5 deleted, suspended or retrieval limit exceeding users, the final dataset consisted of 24 distinctive.

Appendix B: Codebook

B.1 Anxiety Lexicon Codebook:

B.1.1 Direct Anxiety Expressions

Direct anxiety refers to explicit linguistic expressions of fear, risk perception, or perceived harm related to data use, consent, and AI training. These terms signal heightened emotional response and explicit concern.

| Category | Subtype | Example Terms |
| --- | --- | --- |
| Direct Anxiety | Risk perception | risk, risky, dangerous, unsafe, threat |
| Direct Anxiety | Fear & concern | fear, afraid, scared, worried, anxious |
| Direct Anxiety | Harm & violation | harm, damage, violation, abuse |
| Direct Anxiety | Coercion & loss of agency | forced, mandatory, no choice, trapped |
| Direct Anxiety | Exploitation | exploit, scraping, harvesting, extraction |

B.1.2 Metaphorical and Indirect Anxiety Expressions

Metaphorical anxiety captures indirect, figurative, or narrative expressions of loss of control, dispossession, or vulnerability. These expressions often frame privacy harm through metaphor rather than explicit emotional vocabulary.

| Category | Subtype | Example Terms |
| --- | --- | --- |
| Metaphorical Anxiety | Loss of control | losing control, out of control |
| Metaphorical Anxiety | Theft & dispossession | stolen, theft, robbed, taken |
| Metaphorical Anxiety | Surveillance & exposure | watching, spying, surveillance |
| Metaphorical Anxiety | Data misuse | misused, abused, repurposed |
| Metaphorical Anxiety | Vulnerability | exposed, unprotected, defenseless |

B.2 Value Codebook:

• Privacy:

    *Definition: Discourse concerning the control, flow, and access to personal information.*

    *Context: In our dataset, this includes concerns about data scraping, non-consensual model training, surveillance, and the "right to opt-out" (Greene & Shilton, 2018).*

• Intellectual Property (IP):

    *Definition: Discourse focused on ownership, copyright, and the labor of creation.*

    *Context: This captures the specific anxiety of the "Sensitive Group" (artists/writers) regarding their work being used as raw material for AI without compensation or credit. It represents the value of "creative sovereignty."*

• Authenticity:

    *Definition: Justifications related to "realness," truthfulness, and the distinction between human and automated behavior.*

    *Context: Drawing from Hallinan et al. (2022), this includes Twitter's rules against manipulation/spam, but also user demands for "human-only" spaces in the age of AI bots.*

• Community:

    *Definition: References to social connection, belonging, and the network of relationships.*

*Context: As noted by Scharlach et al. (2023), platforms often mobilize "community" to enforce norms. Here, we code for both the preservation of community (users wanting to stay together) and the fracture of community (users mourning the loss of their network).*

- Safety:

    *Definition: Discourse regarding protection from harm, harassment, and toxicity.*

    *Context: This aligns with Twitter's standard "Safety" pillar (e.g., hate speech policies) but also includes user fears of AI-facilitated harassment or deepfakes.*

- Others:

    *Definition: Content unrelated to platform governance values, such as general life updates, sports, or commercial promotion without value-laden commentary.*

Appendix C: Metrics Calculation

C.1 Anxiety Score Construction

First, a lexicon-based anxiety score is computed with custom dictionary (e.g. "no consent", "this feels unsafe", "forced to agree") from the netnography process. The raw lexicon score is normalized to a –1 to +1 range:

$$s_{lex} = 2s_{lex,raw} - 1 \quad (1)$$

Second, a prompt-based anxiety score was calculated using the GPT-4o-mini model between –1(no anxiety) and +1(high anxiety). This model was instructed to establish its judgment on linguistic cues related to fear, concern, uncertainty, and perceived privacy violations. (See Appendix B for detailed prompt)

The final anxiety is then computed by averaging both scores but on more belief in the LLM-generated score using Equation (2), offering a more robust measurement for user anxiety.

$$A_{final} = 0.6s_{gpt} + 0.4s_{lex} \quad (2)$$

For example, a tweet stating *"I feel forced to give up my data and this feels extremely unsafe"* receives a high Anxiety Score due to both explicit anxiety terms and contextual passive emotions. The full workflow is as below:

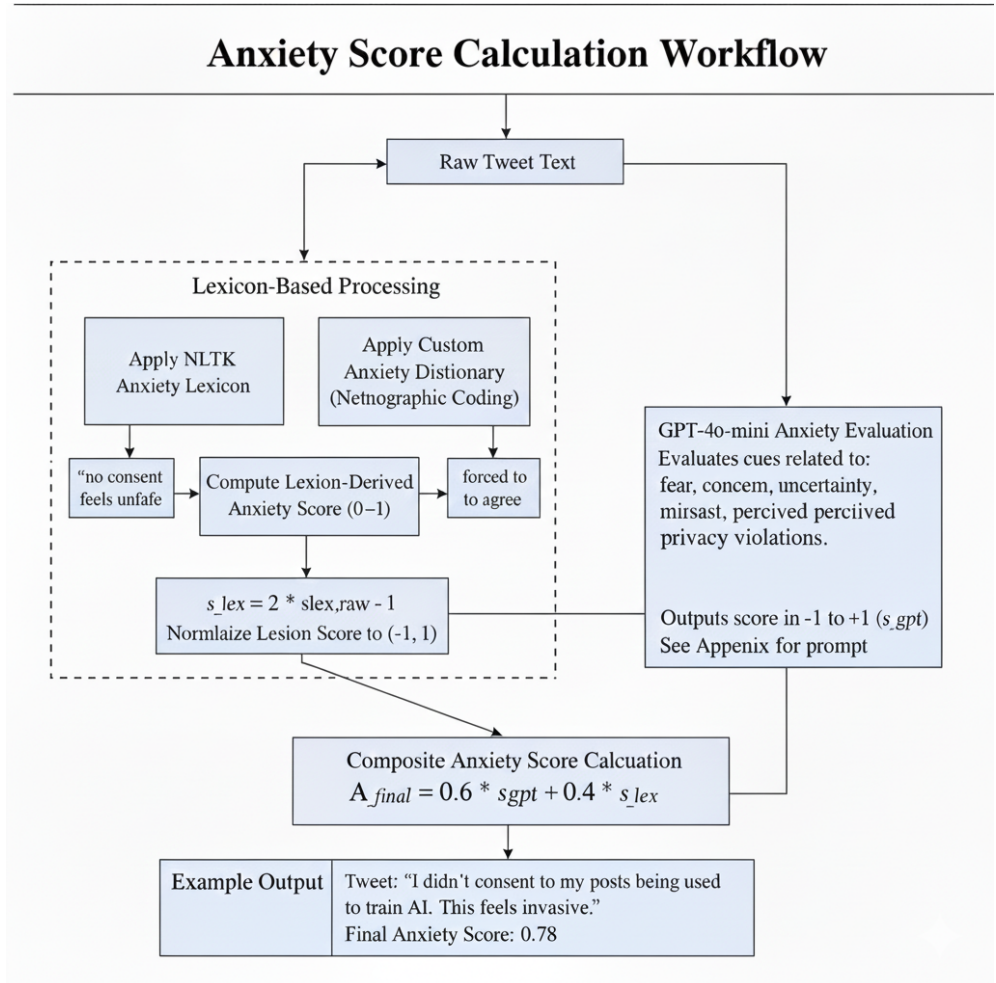

Figure 1: Anxiety Score Calculation

## C.2 Topic Value Construction

Classification results were aggregated at the corpus level to enable comparison of value distributions across communities and time periods. Full technical specifications, including model prompts, value design, and normalization procedures, are provided in the supplementary materials.

We coded platform values at a sentence level to understand how X consider values as a noun (i.e. the specific valuable object) and verb (i.e. the process by which a value is constructed or legitimized) (Kraatz MS, 2020).Seven types of values are defined based on the research purpose and relevant literature, documented in our Taxonomy book(Section 3.2): Privacy, Intellectual Property (IP), Authenticity, Community, Safety, and Others. For Intellectual Property (IP) category specifically, we applied a secondary hierarchical coding layer to determine polarity (Positive/Negative/Neutral), capturing the stakeholders' specific attitude toward ownership rights.

Finally, raw topic counts were normalized by the total number of sentences per corpus, allowing accountability for direct comparisons of value across groups and time.

C.3 Anxiety Score:

Engagement: It is calculated using a simple composite:

$$engagement = reply + quote + agree \quad (3)$$

This metric is calculated at a tweet level and aggregated by community and time.

C.4 Risk Attention Index (RAI):

Higher values indicate more response and greater amplification potential.

$$\mathbf{RAI} = \frac{\#\mathbf{risktweets}}{\#\mathbf{totaltweets}} \quad (4)$$

C.5 Risk-Nonrisk Engagement Ratio (RER):

$$\mathbf{RER} = \frac{\overline{\mathbf{engagement_{risk}}}}{\mathbf{engagement_{non-risk}}} \quad (5)$$

## Appendix D: Prompts

D.1 Value Detection

*You are an assistant for thematic analysis of social media posts from creators, artists, and developers on Twitter (X).*

*Your task is to read the given text and classify it into exactly ONE of the following seven categories:*

*# privacy*

*Use this label if the text discusses personal data, data protection, data collection, user consent, surveillance, tracking, profiling, or control over one's own information. This includes concerns about how the platform stores, shares, or processes personal data, but NOT issues specifically about ownership of creative works or copyright (those belong to IP-\* labels).*

*# authenticity*

*Use this label if the text is about truthfulness, deception, impersonation, fake accounts, bots, spam, misleading content, synthetic or AI-generated media that may confuse users, or requirements to disclose when content is not "real" or authored by the person claiming it.*

*# community*

*Use this label if the text focuses on social groups, inclusion or exclusion of certain people, community norms, shared spaces, belonging, mutual respect, or how the platform frames "our community" or "the public conversation." This includes valuing or devaluing certain groups, or talking about who is welcome or unwelcome on the*

*platform.*

*# IP-positive*

*Use this label if the text talks about protecting creators' rights, enforcing copyright, respecting authorship, giving credit, or safeguarding ownership of user-generated content. Typical examples include policies against copyright infringement, DMCA processes, or statements that users retain ownership of what they create.*

*# IP-negative*

*Use this label if the text talks about stealing, reusing, extracting, or exploiting user-generated content, or the new ToS which reduce user control or ownership for AI model training. This includes clauses that say the platform may use, reproduce, modify, or train AI on user content, or complaints that the platform is "stealing" or "scraping" users' work.*

*# others*

*Use this label if the text does not fit any of the categories above.*

*Important rules:*

*You must output ONLY one single word: privacy, community, safety, IP-positive, IP-negative, IP-neutral, authenticity or others.*

*Do not provide explanations, reasoning, or additional text.*

*Choose the best matching category based on the content. Try your best fit the text into the first six categories except others.*

*Text to analyze:{data}*

D.2 Anxiety Score

You are an assistant that evaluates the level of anxiety expressed in a piece of text.

*Your task is to read the given text and output one single floating-point number representing the anxiety score.*

*# Scoring rules:*

*## The score must be between -1 and 1.*

*1.00 = extremely anxious*

*-1.00 = not anxious at all*

*0.00 = neutral or unclear anxiety*

*Output a value such as: 0.27, -0.53, 0.91, etc.*

*# Output format requirements:*

*Output only one number, nothing else*

*# Format: floating-point, two decimal places*

*No explanation, no description, no text besides the number*

*{data}*

BIBLIOGRAPHY

Acquisti, A. B. (2015). Privacy and human behavior in the age of information. *Science, 347*(6221), 509–514. https://doi.org/10.1126/science.aaa1465.

Berger, J., & Milkman, K. L. (2012). What Makes Online Content Viral? *Journal of Marketing Research, 49*(2), 192–205. https://doi.org/10.1509/jmr.10.0353

Borning, A., & Muller, M. (2012). Next steps for value sensitive design. In *Proceedings of the SIGCHI Conference on Human Factors in Computing Systems (CHI '12)* (pp. 1125–1134). Association for Computing Machinery. https://doi.org/10.1145/2207676.2208560

Chan, N. K., Su, C. C., & Shore, A. (2023). Shifting platform values in community guidelines: Examining the evolution of TikTok's governance frameworks. *New Media & Society, 27*(2), 1127–1151. https://doi.org/10.1177/14614448231189476

Davies, S. R., Wells, R., Zollo, F. and Roche, J. (2024). Unpacking social media `engagement': a practice theory approach to science on social media. *JCOM: Journal of Science Communication, 23*(06), Y02. https://doi.org/10.22323/2.23060402

Fiesler, C., & Hallinan, B. (2018). "We are the product": Public reactions to online data sharing and privacy controversies in the media. In *Proceedings of the 2018 CHI Conference on Human Factors in Computing Systems (CHI '18)* (Paper 53, pp. 1–13). Association for Computing Machinery. https://doi.org/10.1145/3173574.3173627

Fiesler, C., Lampe, C., and Bruckman, A.S. (2016). Reality and Perception of Copyright Terms of Service for Online Creative Communities. In *CSCW '16: Proceedings of the 19th ACM Conference on Computer-Supported Cooperative Work & Social Computing*, (pp. 1450–1461). Association for Computing Machinery. https://doi.org/10.1145/2818048.2819931

Granovetter, M. S. (1973). The Strength of Weak Ties. *American Journal of Sociology, 78*(6), 1360–1380. http://www.jstor.org/stable/2776392

Hallinan, B., Scharlach, R., & Shifman, L. (2022). Beyond Neutrality: Conceptualizing Platform Values. *Communication Theory, 32*(2), 201–222. https://doi.org/10.1093/ct/qtab008

Jeong, U., Nirmal, A., Jha, K., Tang, S. X., Bernard, H. R., & Liu, H. (2024). User Migration across Multiple Social Media Platforms. In S. Shekhar, V. Papalexakis, J. Gao, Z. Jiang, & M. Riondato (Eds.), *Proceedings of the 2024 SIAM*

*International Conference on Data Mining, SDM 2024* (pp. 436–444). Society for Industrial and Applied Mathematics Publications.

Kasperson, R.E., Renn, O., Slovic, P., Brown, H.S., Emel, J., Goble, R., Kasperson, J.X. and Ratick, S. (1988), The Social Amplification of Risk: A Conceptual Framework. *Risk Analysis*, 8, 177–187. https://doi.org/10.1111/j.1539-6924.1988.tb01168.x

Keskin, Batuhan. (2018). Van Dijk, Poell, and de Wall, The Platform Society: Public Values in a Connective World (2018). *Markets, Globalization & Development Review. 3*(3), article 8. 10.23860/MGDR-2018-03-03-08

Kozinets, R.V. (2010) *Netnography: Doing Ethnographic Research Online*. Sage Publications, London.

Kraatz, M. S., Flores, R., & Chandler, D. (2020). The value of values for institutional analysis. *Academy of Management Annals, 14*(2), 474–512. https://doi.org/10.5465/annals.2018.0074

Kraatz, M.S., & Block, E.S. (2008). *Organizational Implications of Institutional Pluralism.*

Kramer, A. D., Guillory, J. E., & Hancock, J. T. (2014). Experimental evidence of massive-scale emotional contagion through social networks. *Proceedings of the National Academy of Sciences of the United States of America, 111*(24), 8788–8790. https://doi.org/10.1073/pnas.1320040111

Kristof, A. L. (1996). Person-organization fit: An integrative review of its conceptualizations, measurement, and implications. *Personnel Psychology, 49*(1), 1–49. https://doi.org/10.1111/j.1744-6570.1996.tb01790.x

Pidgeon, N., Kasperson, R., & Slovic, P. (Eds.). (2003). *The social amplification of risk*. London: Cambridge University Press.

Shilton, Katie. (2013). Values Levers: Building Ethics Into Design. S*cience, Technology, & Human Values, 38,* 374–397. 10.2307/23474474.

Stieglitz, S., & Dang-Xuan, L. (2013). Emotions and Information Diffusion in Social Media—Sentiment of Microblogs and Sharing Behavior. *Journal of Management Information Systems, 29*, 217 – 248.

Stutzman, Fred, Ralph Gross, and Alessandro Acquisti. 2013. "Silent Listeners: The Evolution of Privacy and Disclosure on Facebook". *Journal of Privacy and Confidentiality 4*(2). https://doi.org/10.29012/jpc.v4i2.620

Teigen, K. H. (1994). Yerkes-Dodson: A law for all seasons. *Theory & Psychology, 4*(4), 525–547. https://doi.org/10.1177/0959354394044004

Tufekci, Z. (2015). Algorithmic Harms beyond Facebook and Google: Emergent Challenges of Computational Agency. *Colorado Technology Law Journal, 13*, 203–218.

van Dijck, Jose, The Culture of Connectivity: A Critical History of Social Media (New York, 2013; online edn, Oxford Academic, 24 Jan. 2013), https://doi.org/10.1093/acprof:oso/9780199970773.001.0001

Vosoughi, S., Roy, D., and Aral, S. (2018) ,The spread of true and false news online. *Science, 359*, 1146–1151. DOI:10.1126/science.aap9559

X. (n.d.). *Intellectual property.* Retrieved from X: https://help.x.com/en/rules-and-policies#intellectual-property

X. (n.d.). *General rules.* Retrieved from X.: https://help.x.com/en/rules-and-policies#general

Yerkes, R.M. and Dodson, J.D. (1908), The relation of strength of stimulus to rapidity of habit-formation. *Journal of Comparative Neurology and Psychology, 18*(5), 459–482. https://doi.org/10.1002/cne.920180503